\documentclass[journal]{IEEEtran}
\usepackage{xcolor}
\usepackage{gensymb}

\usepackage{capekCommands}

\usepackage[]{hyperref}
\hypersetup{
    colorlinks,
    linkcolor={red!50!black}, 
    citecolor={blue!50!black},
    urlcolor={blue!20!black}
}

\usepackage{graphicx}

\ifCLASSOPTIONcompsoc
  \usepackage[caption=false,font=normalsize,labelfont=sf,textfont=sf]{subfig}
\else
  \usepackage[caption=false,font=footnotesize]{subfig}
\fi

\newacro{VSWE}{vector spherical wave expansion}
\newacro{VSW}{vector spherical wave}
\newacro{QCQP}{quadratically constrained quadratic program}
\newacro{AToM}{Antenna Toolbox for Matlab~\cite{atom}}
\newacro{FEKO}{FEKO~\cite{feko}}
\newacro{CST}{CST~\cite{cst}}
\newacro{FDTD}{finite diferences in the time domain}

\begin{document}

\pagestyle{headings}
\twocolumn

\title{
Maximum Radiation Efficiency of Arbitrarily-Shaped Implantable Antennas
}
\author{Jakub~Liska,
        Mingxiang~Gao,
        Lukas~Jelinek,
        Erik~R.~Algarp,
        Anja~K.~Skrivervik,
        Miloslav~Capek,~\IEEEmembership{Senior~Member,~IEEE}
\thanks{Manuscript received \today; revised \today.}
\thanks{This work was supported by the Czech Science Foundation under project~\mbox{No.~21-19025M}, and by the Grant Agency of the Czech Technical University in Prague under project \mbox{No.~SGS22/162/OHK3/3T/13}.}
\thanks{J.~Liska, L. Jelinek and M. Capek are with the Czech Technical University in Prague, Prague, Czech Republic (e-mails: \{jakub.liska; lukas.jelinek; miloslav.capek\}@fel.cvut.cz).}
\thanks{M.~Gao, E.~Algarp and A.~K.~Skrivervik are with École Polytechnique Fédérale de Lausanne, Lausanne, Switzerland (e-mails: \{mingxiang.gao; erik.algarp; anja.skrivervik\}@epfl.ch).}
\thanks{Color versions of one or more of the figures in this paper are
available online at http://ieeexplore.ieee.org.}
}

\maketitle

\begin{abstract}
Performance limitations for implanted antennas, taking radiation efficiency as the metric, are presented. The performance limitations use a convex optimization procedure with the current density inside the implant acting as its degree of freedom. The knowledge of the limitations provides useful information in design procedure and physical insight. Ohmic losses in the antenna and surrounding tissue are both considered and quantitatively compared. The interaction of all parts of the system is taken into account in a full-wave manner via the hybrid computation method. The optimization framework is thoroughly tested on a realistic implanted antenna design that is treated both experimentally and as a model in a commercial electromagnetic solver. Good agreement is reported. To demonstrate the feasibility of developed performance limitations, they are compared to the performance of a loop and a dipole antenna showing the importance of various loss mechanisms during the design process. The trade-off between tissue loss and antenna ohmic loss indicates critical points at which the optimal solution drastically changes and the chosen topology for a specific design should be changed.
\end{abstract}


%
\IEEEpeerreviewmaketitle

\section{Introduction}
%
%
%
%


\IEEEPARstart{T}{elemedicine} has experienced a significant boom during the past few decades and important enabler of this progress is wireless communication with, and wireless power transfer to, medical implants via radiofrequency fields, which renders the antenna a crucial element. Laws of physics, however, bound the performance of antennas, and the knowledge of these limits is essential to improve the development of telemedicine devices.

Performance limitations on antennas are of key interest to the antenna designers~\cite{Chu_PhysicalLimitationsOfOmniDirectAntennas, Wheeler_FundamentalLimitationsOfSmallAntennas,Fano_TheoreticalLimitationsOnTheBroadbandMatchingOfArbitraryImpedances,McLean_AReExaminationOfTheFundamentalLimitsOnTheRadiationQofESA,YaghjianBest_ImpedanceBandwidthAndQOfAntennas,Harrington_OnTheGainAndBWofDirectAntennas} even outside telemedicine and their determination has evolved to a discipline on its own. For antennas radiating into free space, they can nowadays be evaluated for arbitrary metrics and shapes~\cite{Gustafsson_OptimalAntennaCurrentsForQsuperdirectivityAndRP,2020_Molesky_PRR,GustafssonCapek_MaximumGainEffAreaAndDirectivity}.
Nevertheless, the aspects of wireless communication with implanted devices~\cite{wearableImplantableAntennasReview,Nikolayev_AntennaIngestibleAndImplantableApp} differ significantly from antennas operating in free space. Of particular interest for this paper are the performance limitations on the maximum power density of \ac{EM} waves reaching free space from implanted antennas~\cite{Skrivervik_Bosiljevac_Sipus_FundamentalBoundsForImplantedAntennas,Bosiljevac_PropagationInFiniteLossyMedia,Gao_RulesOfThumbToAssessLossesOfImplantedAntennasEuCAP21,Gao_OnTheMaximumPowerDensityofImplantedAntennasWithinSimplifiedBodyPhantomsEuCAP22, Nikolayev+etal2019}. Generally, the size of an implanted antenna is typically a few millimeters, and the operating frequency ranges from hundreds of megahertz to a few gigahertz~\cite{Nikolayev2018}. This indicates that implanted antennas are usually electrically small. Contrary to conventional antennas, the implant is firstly surrounded by a small volume of ideally lossless medium representing the implant encapsulation. Beyond this, the \ac{EM} waves radiated by the implant must pass through a lossy tissue before escaping to free space. As such, performance limitations determine the benchmark for every human or computer powered design~\cite{2020_Molesky_PRR,Liska-CompFunBoAntennas-EuCAP22}.

Until recently, performance limitations for implanted antennas have only been treated considering elementary sources~\cite{Skrivervik_Bosiljevac_Sipus_FundamentalBoundsForImplantedAntennas} with no considerations on true physical dimensions. A finite geometry and a cylindrical shell were considered in~\cite{Nikolayev+etal2019} and compared to infinitesimally small sources studied before. An initial attempt to find the radiation efficiency limitation of an implanted antenna with finite conductivity and arbitrary shape of the current-supporting region was presented in~\cite{Jelinek-MaxRadEffImplantedAntennaHybrid, Liska_MaxRadEffImplantableAntennaEuCAP2023}. It was followed by a feasibility study of performance limitations on implanted antennas considering arbitrarily-shaped and lossy current-supporting regions~\cite{Algarp2022}.

\begin{figure}
    \centering
    \includegraphics{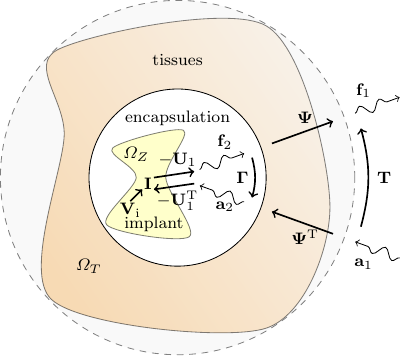}
    \caption{Within the used hybrid method, the model is separated into regions treated by different numerical methods: \ac{MoM} description of the current-supporting region denoted as $\varOmega_Z$, and \ac{VSWE} description of the host body denoted as $\varOmega_T$. The spherical boundary around the antenna (implanted capsule) separates the two methods and provides their interconnection. Current density in the current-supporting region $\varOmega_Z$ is described via vector~$\Ivec$, while \ac{VSWE} is used on the boundaries of the host body $\varOmega_T$ with expansion coefficients collected in vectors $\M{a}_1, \ \M{f}_1$ and $\M{a}_2, \ \M{f}_2$ outside and inside the host body, respectively. The coupling of both methods is represented by matrices~$-\Umat_1$ and~$-\Umat_1^\trans$. Field interaction with $\varOmega_T$ region is given by $\M{\Gamma}, \ \M{T}, \ \M{\Psi}, \ \M{\Psi}^\trans$ matrices.}
    \label{fig:setupLayer}
\end{figure}

A popular methodology for evaluating performance limitations is to find the optimal solution for a specific scenario. For instance, the optimal current density within an allowed volume can be obtained using an \ac{EFIE} formulation combined with a quadratic optimization scheme. This has, however, the disadvantage that in the case of implanted devices, the \ac{MoM} formulation used to solve the \ac{EFIE} will require a large number\footnote{A direct approach using the volumetric meshing of inhomogeneous tissues would be unbearably computationally demanding~\cite{Losenicky_etal_MoMandThybrid}.} of unknowns to account for the interaction of the electrically small implant with a potentially large host body\footnote{The radius of the smallest sphere circumscribing the body is much larger than a typical operating wavelength in the free space.}.
To overcome the computation difficulties, a hybrid method~\cite{Kim-ImplantedAntennasInsideHumanBody} is used in this work, which consists of a combination of two frequency domain schemes: the \ac{MoM} formulation of the \ac{EFIE} is used for the antenna and its closest neighborhood, while a \ac{VSWE} is applied to numerically evaluate the interaction of the \ac{EM} field with the host body~\cite{2022_Losenicky_TAP}. This allows treating this complex problem with relatively few unknowns~\cite{2022_Losenicky_TAP}. The hybrid method requires a spherical boundary on which the two underlying numerical methods, \ac{MoM} and \ac{VSWE}, are coupled. This boundary is defined naturally as a spherical encapsulation of the current support, see~Fig.~\ref{fig:setupLayer}.

The aim of this work is to use the hybrid method as formalized in \cite{2022_Losenicky_TAP} to create the framework to evaluate fundamental limitations on the efficiency of implanted antennas, considering internal antenna losses and the losses in the host body. The performance potential of the design region, the idea of which was introduced in~\cite{Jelinek-MaxRadEffImplantedAntennaHybrid, Algarp2022, Liska_MaxRadEffImplantableAntennaEuCAP2023}, can be assessed before starting an antenna design for a specific problem. This manuscript comprehensively explores the quest for optimal radiation efficiency. Profound insights are provided through a detailed analysis of the physical operations and the impact of various phenomena. This elucidates a clear understanding of why the optimal current density takes specific shapes and why these shapes exhibit significant variations, even in relatively similar scenarios.

The framework employs previously developed optimization schemes~\cite{Liska_etal_FundamentalBoundsEvaluation,Liska-CompFunBoAntennas-EuCAP22}. As with antennas in the free space~\cite{Gustafsson_OptimalAntennaCurrentsForQsuperdirectivityAndRP, Liska_etal_FundamentalBoundsEvaluation}, the computation of performance limitations~\cite{Liska_etal_FundamentalBoundsEvaluation,Liska-CompFunBoAntennas-EuCAP22} is based on convex optimization, particularly~\ac{QCQP}~\cite{NocedalWright_NumericalOptimization,Liska_etal_FundamentalBoundsEvaluation}. The approach is applied to a basic scenario with a spherical single-layered phantom and commonly used medical frequencies of 403\,MHz and 2.45\,GHz. Performance limitations show the contribution of different regions to the dissipation and shape of the optimal current density which can inspire the design. The feasibility of performance limitations is also analyzed, resulting in further design recommendations. Results evaluated in AToM package~\cite{atom} are verified by measurement and by simulations using independent commercially available solvers: \ac{CST} simulation based on \ac{FDTD} and \ac{FEKO} simulation based on \ac{MoM}.

This paper is organized as follows. Section~\ref{sec:formulation} presents the used model and the basis of the numerical tools used later. Their description is necessary to understand the following sections. Section~\ref{sec:losses} stresses the importance of finite metal conductivity of the current-supporting region and consequences. The method verification is done in Section~\ref{sec:verification}, which also presents results of simulations in commercially available solvers and measurement results. Further investigation and discussion in the setup for verification are given in Section~\ref{sec:investigation}.

\section{Computational Tools}\label{sec:formulation}
In the context of maximizing radiation efficiency, this section provides the necessary theoretical foundation for determining optimal current density distribution within the designated design region.

Modeling the wireless link between an implanted antenna and an external node is a challenging problem, resolved here using the advantages of two distinct approaches, \ac{MoM} and \ac{VSWE}, which are combined within the hybrid method~\cite{2022_Losenicky_TAP}. \Ac{MoM} with piece-wise basis functions~\cite{RaoWiltonGlisson_ElectromagneticScatteringBySurfacesOfArbitraryShape} is applied to characterize the current-supporting region\footnote{Allowed design region, support for the optimized current density in the sense of the performance limitations defined in Section~\ref{sec:formulation}, or simply the antenna metalization in the case of a design.} (region~$\varOmega_Z$ in Fig.~\ref{fig:setupLayer}) in a great level of spatial details. Conversely, \ac{VSWE} serves as a set of entire-domain basis functions to determine the influence of the surrounding tissues (region~$\varOmega_T$ in Fig.~\ref{fig:setupLayer}). Computational complexity is significantly reduced~\cite{Moerlein2021AntennaDe-EmbFDTDusinfSWFarXiv,Berkelmann2022AntennaOptimWBANbasedSWFde-EmbTAP} as compared to the case, in which the entire setup is solved with a single method with entire-domain discretization. The hybrid method used in this work is briefly recapitulated in Appendix~\ref{app:hybrid}, and computation complexity is discussed in Appendix~\ref{app:computation}. The appendices also briefly discuss the methodological limitations of the hybrid approach.

The hybrid method allows a matrix system
describing the entire scenario to be set up, including both the antenna and host body. Based on equations defining \ac{MoM}, \ac{VSWE}, and their interconnection, the description reads
\begin{equation}\label{eq:system}
    \begin{bmatrix}
        \Zmat_0 + \Zmat_\rho & - \M{U}_1^\trans & \M{0} & \M{0} \\
        -\M{U}_1  & \M{0} & -\M{1} & \M{0} \\
        \M{0}  & -\M{1} & \M{\Gamma} & \M{0} \\
        \M{0}  & \M{0} & \M{\Psi} & -\M{1}
    \end{bmatrix}  \begin{bmatrix}
        \Ivec \\ \M{a}_2 \\ \M{f}_2 \\ \M{f}_1
    \end{bmatrix} = \begin{bmatrix}
        \Vvec_\T{i} \\ \M{0} \\ \M{0}\\ \M{0}
    \end{bmatrix},
\end{equation}
where identity and zero matrices are denoted by~$\M{1}$ and $\M{0}$, respectively. The meaning of the individual rows is illustrated in Fig.~\ref{fig:setupLayer} and reads as follows:
\begin{enumerate}
    \item Current~$\M{I}$ in region~$\varOmega_Z$ results from a reaction of impedance matrix $\Zmat_0 + \Zmat_\rho$ on excitation, which is given by direct excitation~$\Vvec_\T{i}$ and indirect excitation~$\M{U}_1^\trans \M{a}_2$ resulting from the reflected field~$\M{a}_2$, see Fig.~\ref{fig:setupLayer}.
    \item Outgoing \acp{VSW}~$\M{f}_2$ are generated via current vector~$\M{I}$ through the coupling matrix~$\Umat_1$.
    \item Incoming \acp{VSW}~$\M{a}_2$ inside the encapsulation are generated by reflected outgoing \acp{VSW}~$\M{f}_2$ via matrix~$\M{\Gamma}$.
    \item Outgoing \acp{VSW}~$\M{f}_1$ outside the host body result from penetrating outgoing \acp{VSW}~$\M{f}_2$. The phenomenon is expressed by matrix~$\M{\Psi}$.
\end{enumerate}
For further details about the hybrid method and the matrices, see Appendix~\ref{app:hybrid} and~\cite{2022_Losenicky_TAP}.

Within the quadratic representation of power-like quantities in \ac{MoM} and \ac{VSWE}, the performance limitations are readily evaluated as in previous studies~\cite{Gustafsson_OptimalAntennaCurrentsForQsuperdirectivityAndRP,Liska-CompFunBoAntennas-EuCAP22,Liska_etal_FundamentalBoundsEvaluation}. The evaluation yields the value of the fundamental bound on a chosen metric and an associated optimal current distribution which may serve as the initial inspiration for antenna designers.

Concerning implanted antennas, radiation efficiency is one of the most important performance metrics. The performance is generally low due to the lossy tissues surrounding the antenna. The previously studied limits~\cite{Skrivervik_Bosiljevac_Sipus_FundamentalBoundsForImplantedAntennas, Gao_RulesOfThumbToAssessLossesOfImplantedAntennasEuCAP21} considered cycle-mean radiated power $P_\T{rad}$ and cycle-mean power absorbed by the lossy biological tissues $P_\T{tis}$ due to near-field losses and field propagation through the host body.

Within the hybrid method shown in Appendix~\ref{app:hybrid}, ohmic losses in the antenna can also be considered and separated from the total loss. Similarly to~\cite{Jelinek-MaxRadEffImplantedAntennaHybrid, Algarp2022, Liska_MaxRadEffImplantableAntennaEuCAP2023}, the upper bound on radiation is here formulated as the optimal current density, which realizes the highest radiation efficiency. The expansion coefficients~$I_n$ are taken as the degrees of freedom in this optimization problem. Explicitly, the radiated power~$P_\T{rad}$ over the total power~$P_\T{tot}$ supplied to the system is maximized
\begin{equation}\label{eq:optim}
    \eta_\T{rad}^\T{ub} = \max \limits_\Ivec \eta_\T{rad},
\end{equation}
where
\begin{equation}
     \eta_\T{rad} = \dfrac{P_\T{rad}}{P_\T{tot}} = \dfrac{P_\T{rad}}{P_\T{rad} + P_\T{ant} + P_\T{tis}},
\end{equation}
with~$P_\T{ant}$ being the cycle-mean lost power in the current-supporting region~$\varOmega_Z$, and $P_\T{tis}$ being the cycle-mean lost power in region~$\varOmega_T$. Both powers are assumed to cover conduction and polarization losses as well.

The individual power terms are defined within the hybrid method as follows
\begin{align}
    P_\T{rad} &= \dfrac{1}{2} \M{f}_1^\herm \M{f}_1, \label{eq:Prad} \\
    P_\T{ant} &= \dfrac{1}{2} \Ivec^\herm \RE \left[ \Zmat_\rho \right] \Ivec, \label{eq:Pant} \\
    P_\T{tis} &= \dfrac{1}{2} \left(\M{f}_2^\herm \M{f}_2 + \RE \left[ \M{a}_2^\herm \M{f}_2 \right] \right) - P_\T{rad}. \label{eq:Ptis}
\end{align}
Note that the first term on the right-hand side of~\eqref{eq:Ptis} is the net cycle-mean outward power flow at the inner boundary of~$\varOmega_T$, see~\cite[Appendix~E]{2022_Losenicky_TAP} for derivation. 

As mentioned above, the current density expansion coefficients~$I_n$ are the optimization variables, and the first row of system~\eqref{eq:system}, $\left( \Zmat_0 + \Zmat_\rho \right) \Ivec - \M{U}_1^\trans \M{a}_2 = \Vvec_\T{i}$, is therefore not imposed\footnote{Note that equation system~\eqref{eq:system} has a unique solution, which binds the excitation vector with current vector. In order to set up an upper limit to radiation efficiency of all antennas fitting the prescribed current-supporting region, this unicity must be relaxed. One of the possible ways is to skip the first row of equation system~\eqref{eq:system}, which gives the necessary freedom to the current vector~$\Ivec$. But, particular lines in the first row can be used as power constraints, to further tighten the bounds and to impose constraints on input impedance, the shape of the feed, \textit{etc}.}~\cite{JelinekCapek_OptimalCurrentsOnArbitrarilyShapedSurfaces}. Such relaxation allows a setup of field configuration, the performance of which cannot be overcome by any design since that would satisfy the first row of system~\eqref{eq:system}. The relaxed system only relies on the boundary conditions between different regions, which are represented by other rows. Specifically, the equation system
\begin{equation}\label{eq:linCon}
    \begin{bmatrix}
        -\M{U}_1  & \M{0} & -\M{1} & \M{0} \\
        \M{0}  & -\M{1} & \M{\Gamma} & \M{0} \\
        \M{0}  & \M{0} & \M{\Psi} & -\M{1}
    \end{bmatrix}  \begin{bmatrix}
        \Ivec \\ \M{a}_2 \\ \M{f}_2 \\ \M{f}_1
    \end{bmatrix} = \begin{bmatrix}
        \M{0} \\ \M{0} \\ \M{0}
    \end{bmatrix}
\end{equation}
is used as an affine constraint in the optimization problem, while the current density vector~$\Ivec$ is optimized to maximize radiation efficiency. The explicit solution to the optimization problem is presented in Appendix~\ref{app:optim}, and the computational complexity of the optimization is presented in Appendix~\ref{app:computation}.

\section{Effect of Different Types of Losses}\label{sec:losses}
Implanted antennas suffer from power dissipated in two different regions: the antenna itself and the host body. So far, only idealized \ac{EM} dipole sources were considered in~\cite{Skrivervik_Bosiljevac_Sipus_FundamentalBoundsForImplantedAntennas, Nikolayev+etal2019}, showing that elementary electric dipole suffers more from tissue losses than the magnetic source, mainly due to the coupling of the near field and the lossy tissues, and concluding that the latter should thus be preferred. In this section, we show that the metal losses of a given current-supporting region should also be considered, and may influence the choice of the best type of antenna.

\begin{figure}
    \centering
    \includegraphics{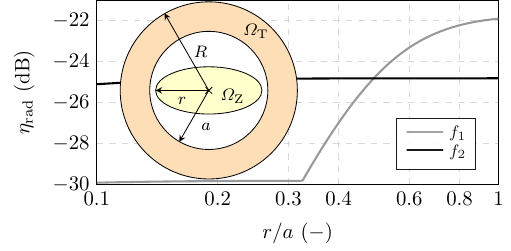}
    \caption{Upper bound on radiation efficiency for frequencies $f_1,f_2$ and radius of the disc normalized by the radius of the air bobble $r/a \in [0.1,1]$. Performance bounds are evaluated for the copper disc as the current support centered inside air bobble with radius $a = 7.5\,\T{mm}$, which is placed in the center of the muscle-filled sphere with radius $R = 61.5\,\T{mm}$.}
    \label{fig:effMuscle}
\end{figure}

Let us consider two frequencies within standard medical bands: $f_1 = 403\,\T{MHz}$ and $f_2 = 2.45\,\T{GHz}$. Further, let us consider the following example: the current-supporting region is a copper disc with radius~$r$ swept from 0.75\,mm to 7.5\,mm placed in a centered spherical air bubble with radius $a = 7.5\,\T{mm}$ and implanted in spherical muscle ($\varepsilon_\T{r,m} = 57 - 36\J$ at frequency~$f_1$ and~$\varepsilon_\T{r,m} = 53 - 13\J$ at frequency~$f_2$) phantom with radius $R = 61.5\,\T{mm}$. The current-supporting region is supposed to be made of copper with a thickness much higher than skin depth and, therefore, can be modeled by a sheet of surface resistivity~\cite{Pozar_MicrowaveEngineering}
$R_\T{s} = \sqrt{(2 \pi f \mu_0)/(2 \sigma)}$,
where conductivity~$\sigma = 5.56 \cdot 10^7 \, \T{S/m}$ and permeability of vacuum~$\mu_0$ are assumed. The upper bound on radiation efficiency is computed for different radii $r$ and is shown in Fig.~\ref{fig:effMuscle}.

Fig.~\ref{fig:effMuscle} depicts, at each frequency, the optimal radiation efficiency as a function of normalized radius~$r/a$. This efficiency cannot be overcome\footnote{Optimal current can be arbitrary, and the optimization routine chooses the best one according to the optimized metric. No antenna built in the allowed design region can perform better than the best overall current density distributions.} by any real antenna fitting the considered disc region within this scenario.

An interesting phenomenon is an abrupt growth in radiation efficiency at frequency~$f_1$ for support sizes larger than $r/a=0.32$. For current supports up to a radius of $r/a = 0.5$, it is convenient to use frequency $f_2$ with stable performance among different radii. In contrast, for larger sizes, the first frequency is preferred.

\begin{figure}
    \centering
    \includegraphics{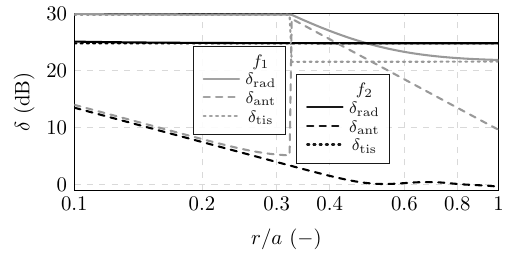}
    \caption{Dissipation factors corresponding to the upper bound on radiation efficiency shown in Fig.~\ref{fig:effMuscle}. The partial dissipation factors of the current-supporting region and the muscle host body are also shown. The radius of the disc is, as in Fig.~\ref{fig:effMuscle}, normalized by the radius of the air bobble.}
    \label{fig:disMuscle}
\end{figure}

A better physical insight into the aforementioned phenomenon is attained by analyzing the power quantities in~\eqref{eq:Prad}--\eqref{eq:Ptis} normalized to the radiated power~$P_\T{rad}$, i.e., dissipation factors of the current support/tissue
\begin{equation}
    \delta_\T{ant/tis} = \dfrac{P_\T{ant/tis}}{P_\T{rad}}, \label{eq:deltaAntTis}
\end{equation}
and the total dissipation factor
\begin{equation}\label{eq:deltaRad}
    \delta_\T{rad} = \delta_\T{ant} + \delta_\T{tis} = \dfrac{1 - \eta_\T{rad}}{\eta_\T{rad}}.
\end{equation}
The comparison of all dissipation factors is shown in Fig.~\ref{fig:disMuscle}.

At frequency~$f_1$, the dissipation factors $\delta_\T{ant}$ and $\delta_\T{tis}$ are discontinuous, with the critical point being the radius $r/a=0.32$. For smaller sizes, tissue loss dominates, while it is comparable to the antenna loss for larger sizes. At frequency~$f_2$, all curves are smooth, and antenna loss is negligible compared to tissue loss.

\begin{figure}
    \centering
     \subfloat{\includegraphics[width = 1.66in]{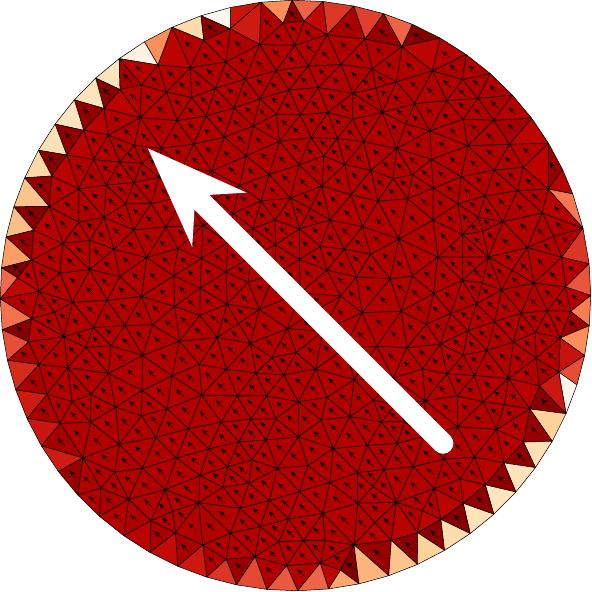}}
     \hfill
    \subfloat{\includegraphics[width = 1.66in]{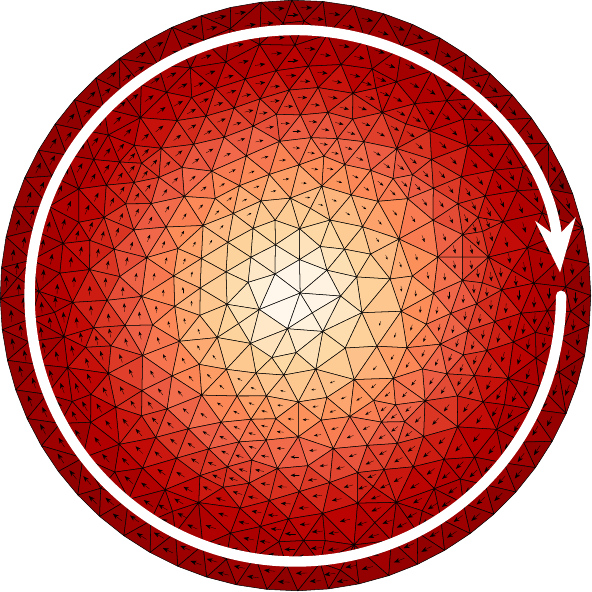}}
     \vfill
      \subfloat{\includegraphics[width = 1.66in]{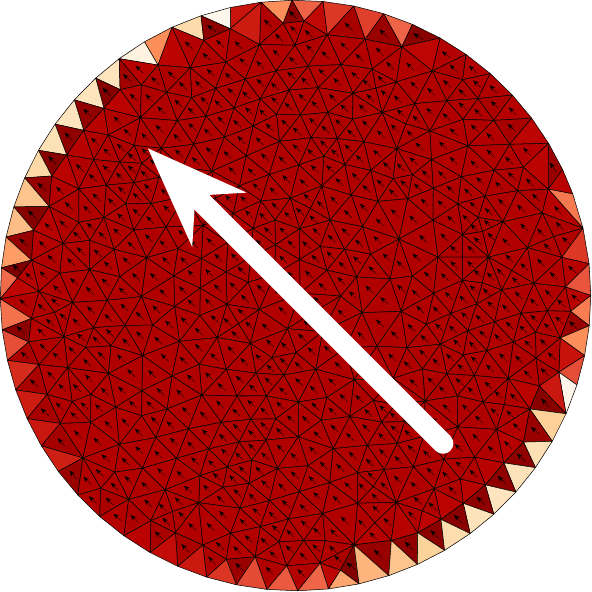}}
     \hfill
      \subfloat{\includegraphics[width = 1.66in]{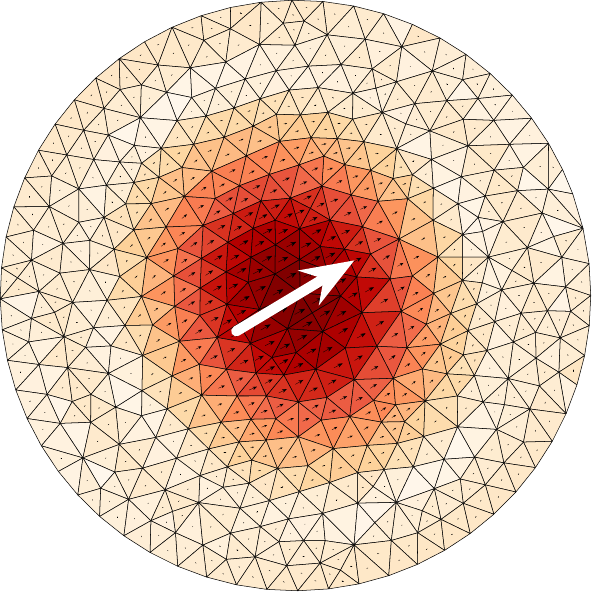}}
     \caption{Optimal current densities on a disc in a muscle host body corresponding to data reported in Fig.~\ref{fig:disMuscle} and Fig.~\ref{fig:effMuscle}: (top left) $\T{TM}_{1m}$-like for $r/a = 0.1$ and $f_1 = 403 \, \T{MHz}$, (top right) $\T{TE}_{1m}$-like for $r/a = 1$ and $f_1 = 403 \, \T{MHz}$, 
 (bottom left) $\T{TM}_{1m}$-like for $r/a = 0.1$ and $f_2 = 2.45 \, \T{GHz}$, (bottom right) $\T{TM}_{1m}$-like for $r/a = 1$ and $f_2 = 2.45 \, \T{GHz}$.}
     \label{fig:optCurrentsMuscle}
\end{figure}

The behavior of power losses is elucidated by examining the optimal current density in scenarios with small and large current-supporting regions. For the smaller sizes, depicted in the left panels of Fig.~\ref{fig:optCurrentsMuscle}, the optimal current is $\T{TM}_{1m}$-like\footnote{The notation designates current density as $\T{TE}_{1m}$-like (magnetic dipole), $\T{TM}_{1m}$-like (electric dipole), or $\T{TM}_{2m}$-like (electric quadrupole). This notation reflects the mode generated when projected onto the \ac{VSWE}: $\T{TE}_{1m}$ mode, $\T{TM}_{1m}$ mode, or $\T{TM}_{2m}$ mode, respectively. The subscript $m$ corresponds to the azimuthal number and reflects a rotation of the coordinate system.}. Conversely, for the larger current-supporting regions, as seen in the right panels of Fig.~\ref{fig:optCurrentsMuscle}, the optimal current density transforms to $\T{TE}_{1m}$-like state at frequency $f_1$ and $\T{TM}_{1m}$-like ibid at frequency $f_2$. This phenomenon accounts for the discontinuity in dissipation factors at frequency $f_1$ where the optimal current undergoes a significant shift in character to maximize radiation efficiency.

The rationale behind this transition lies in the fact that $\T{TE}_{1m}$-like currents exhibit less coupling to the lossy dielectric host body~\cite{Skrivervik_Bosiljevac_Sipus_FundamentalBoundsForImplantedAntennas}. This characteristic becomes advantageous when the borders of the current-supporting region approach the tissue. Analogously, small-size $\T{TE}_{1m}$-like currents can be likened to a short circuit~\cite{Algarp2022}. At frequency $f_2$, this analogy is unnecessary since the electrical size of the current support is sufficient to accommodate a $\T{TM}_{1m}$-like current, achieving an optimal trade-off between antenna loss and tissue loss due to the dense current in the center and the short distance from the tissues.

In general, antenna loss decreases with increasing $r$ for both $\T{TM}_{1m}$-like and $\T{TE}_{1m}$-like currents, as it allows for greater spreading of the current. Regarding computational complexity see Appendix~\ref{app:computation}, where the most demanding scenario: current-supporting region with $r/a = 1$ and frequency $f_1 = 403 \, \T{MHz}$ is discussed.

\section{Experimental Verification}\label{sec:verification}
The following two sections show how to use the previously developed methodology in the real design of an implanted antenna. This provides an experimental validation of the method and also demonstrates some new features that were not observed previously.

\begin{figure}
    \centering
    \includegraphics[scale=0.6]{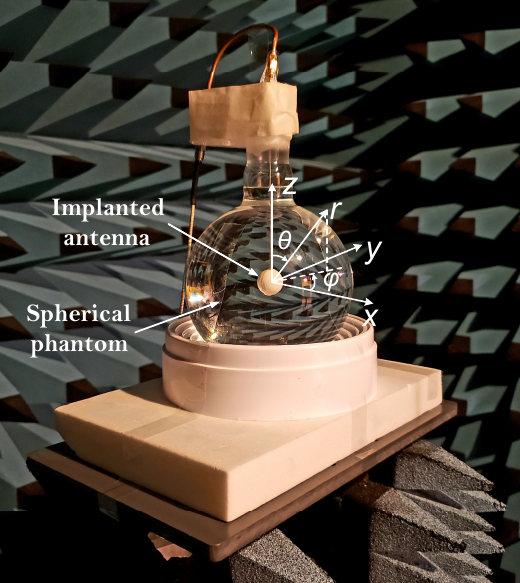}
    \caption{Measurement setup placed in the anechoic chamber. The ceramic-encapsulated antenna is placed in the center of a spherical phantom filled with distilled water.}
    \label{fig:photo}
\end{figure}

\begin{figure}
    \centering
    \includegraphics{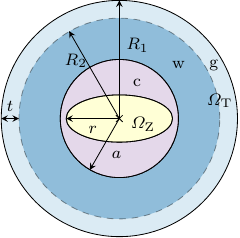}
    \caption{Simplified scenario of the measurement: c -- ceramic, w -- water, g -- glass. The parameters are: spherical glass bottle ($\varepsilon_\T{r,g} = 6.7 $) with an outer radius of $R_1 = 63.5 \, \T{mm}$, a glass thickness $t = 2 \, \T{mm}$, water radius $R_2 = R_1 - t = 61.5 \, \T{mm}$, ceramic encapsulation $a = 7.5 \, \T{mm}$. The true shape of the antenna ($\varOmega_Z$) is detailed in~Fig.~\ref{fig:ant_parameter}. In the case of the fundamental bound, the current-supporting region is a disc of radius $r \in \left[ 0.75, 7.5 \right] \, \T{mm}$.}
    \label{fig:setupWater}
\end{figure}

To allow for a simple experimental setup, the host body is simulated by a glass bottle filled with distilled water, see Fig.~\ref{fig:photo} and Fig.~\ref{fig:setupWater}. Distilled water has well-determined \ac{EM} properties ($\varepsilon_\T{r,w} = 78 - 12\J$ at frequency~$f_2$) and similarly well defined are \ac{EM} properties of glass ($\varepsilon_\T{r,g} \approx 6.7$). These values are also used by commercial \ac{EM} solver \ac{CST}, which is employed as a virtual version of the experiment. The outer radius of the bottle is $R_1 = 63.5\, \T{mm}$ and glass thickness is $t = 2.0\,\T{mm}$.

\begin{figure}
    \centering
    \includegraphics[scale=0.55]{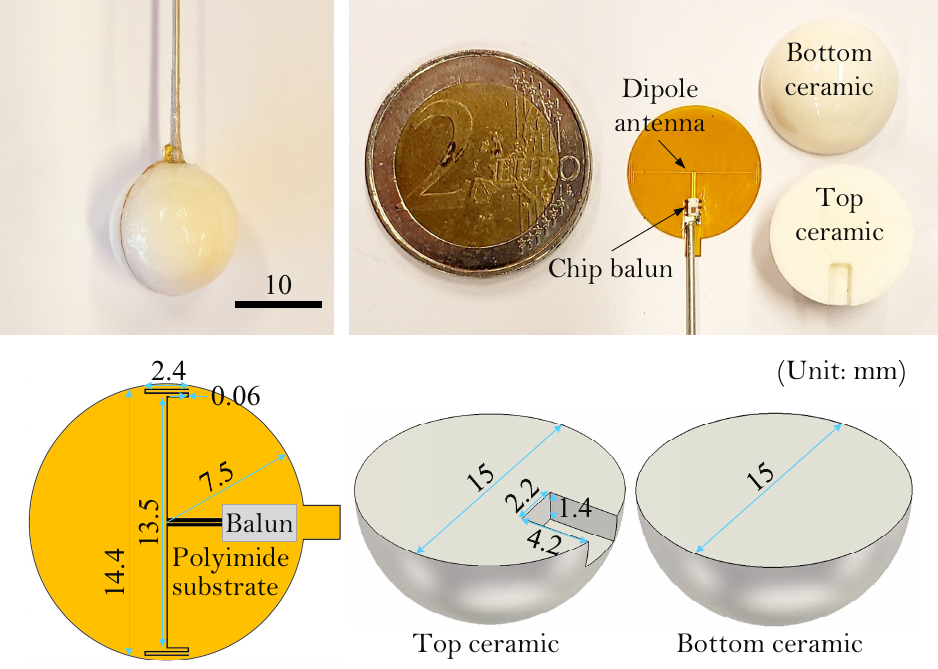}
    \caption{Photograph and design of the ceramic-encapsulated implanted antenna: antenna encapsulated in the ceramic hemispheres (left top panel), decomposed antenna and encapsulation (right top panel), visualization of the meander dipole antenna including dimensions and interconnection with balun via $3.2\,\T{mm}$ long coplanar stripline with a strip width of $0.18\,\T{mm}$ and a gap width of $0.1\,\T{mm}$ (left bottom panel), encapsulation hemisphere's dimensions (center and right bottom panel).}
    \label{fig:ant_parameter}
\end{figure}

The implanted antenna, see Fig.~\ref{fig:ant_parameter}, is designed and measured for frequency $f_2$. It consists of a $100$\,µm thick polyimide disc ($\varepsilon_\T{r} = 3.5$) with $18$\,µm thick copper cladding and two ceramic hemispheres as its encapsulation. A meander dipole antenna is etched in the copper. The antenna is connected to a copper jacketed semi-rigid cable (EZ-34, EZ Form Cable) through a chip balun (2450BL15B050, Johanson Technology) and a short coplanar stripline. A slot in the top ceramic hemisphere is machined to accommodate the balun and the cable. The ceramic hemispheres are made of $\T{Al_2O_3}$ (alumina, $\varepsilon_\T{r} = 12$ and $\T{tan}\delta = 0.0003$, \ac{CST} model) that is bio-compatible and also provides good matching of the antenna.

\begin{figure}
    \centering
    \includegraphics[width = 0.35\textwidth]{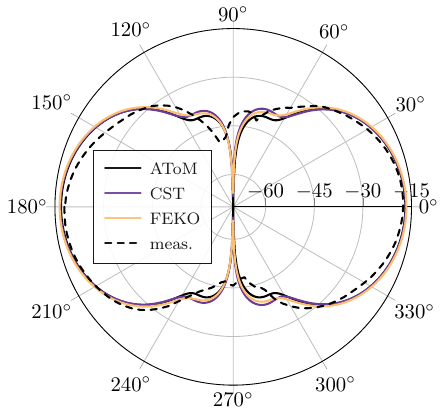}
    \caption{Simulated and measured gain patterns' cuts plotted in dB scale at the frequency $f_2$ using the setup in Fig.~\ref{fig:photo}. The cut plane corresponds to $\theta = 90\degree$. Polarization in the plane of the dipole antenna is measured.}
    \label{fig:gainPattern}
\end{figure}

At first, the above-mentioned antenna setup is used to validate the hybrid method used to define the fundamental bound. For this aim, the gain pattern of the antenna was measured in an anechoic chamber, see Fig.~\ref{fig:gainPattern} and Appendix~\ref{app:meas} for details. Apart from the measurement and the hybrid method, the gain patterns were also evaluated in commercial solvers: \ac{CST} and \ac{FEKO}. All simulations omit the thin, flexible substrate and losses in the ceramic encapsulation. The gain patterns are depicted in Fig.~\ref{fig:gainPattern} and show excellent agreement. The maximum gain values $G(\pi/2, 0)$ are $-16.9 \, \T{dB}$ for simulated pattern by the hybrid method and using \ac{AToM}, $-16.9 \, \T{dB}$ in \ac{CST},  $-16.5 \, \T{dB}$ in \ac{FEKO}, and $-17.6 \, \T{dB}$ for the measurement. The computational requirements of the commercial solutions (using the FDTD method in CST and the surface equivalence formulation in FEKO) were considerable due to the use of high dielectric constants and overall large electric size. The data presented in Fig.~\ref{fig:gainPattern} demanded hours of computer time. Compared to this, the hybrid method demanded only a few seconds and one order in magnitude less computer memory for the same task with the same discretization of the dipole antenna. Even for the maximum possible meshing allowed by the computer used, the results from the commercial solvers were still slightly mesh-dependent, and the hybrid method is therefore considered more reliable for this task. A detailed comparison and discussion on computation requirements is given in Appendix~\ref{app:computation}.

\begin{figure}
    \centering
    \includegraphics{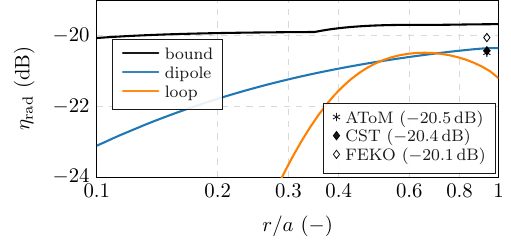}
    \caption{The radiation efficiency of the meandered dipole antenna as simulated with the hybrid method, \ac{CST}, and \ac{FEKO} at frequency $f_2$. The performances of a simple circular loop and a straight dipole antenna are also shown. The fundamental bound on the performance of a disc current-supporting region is the upper bound on the radiation efficiency of all considered antenna designs. The dipole antenna, loop antenna, and fundamental bound correspond to varying sizes of the current support, the size of which is measured by the radius of the smallest circumscribing sphere~$r$. The sizes are normalized to the radius of the ceramic encapsulation, i.e., at ~$r/a = 1$ the antenna fills the entire capsule.}
    \label{fig:effWater}
\end{figure}

The radiation efficiency of the experimental setup, as illustrated in Fig.~\ref{fig:effWater}, undergoes a comparative analysis with loop and dipole antennas. These antennas are constructed from a copper strip with a width of $w = 0.5 \, \T{mm}$ and fed by an ideal delta-gap voltage source. The results exhibit favorable consistency across various numerical simulations, with minor deviations attributed to distinct numerical schemes. Notably, the figure incorporates the fundamental bound on radiation efficiency considering a current-supporting region as a disc that encompasses all utilized antenna models. Computation requirements for the largest current-supporting region are detailed in Appendix~\ref{app:computation}.

The performance of the experimental setup remains approximately~$1\,\T{dB}$ below the fundamental bound. This implies that, given the material composition and shape of the design region (disc), there is limited room for improvement. The hybrid method elucidates that the interaction with the host material is predominantly dipolar, indicating a limited dependence on the specific composition of the host body.

This design is anticipated to perform similarly under real-body conditions. The difference is that the real body is not ideally spherical, and its composition is more complex than pure water. This evidence might cause a slight frequency shift and variance of tissue loss and critical point as well. Similar phenomena may happen once the encapsulation is not ideally lossless and spherical. Furthermore, introducing other conductive components inside the implant can influence the optimal current, overall maximal possible radiation efficiency, and the proportion of power dissipation in different regions.

\section{Discussion}\label{sec:investigation}
\begin{figure}
    \centering
    \includegraphics{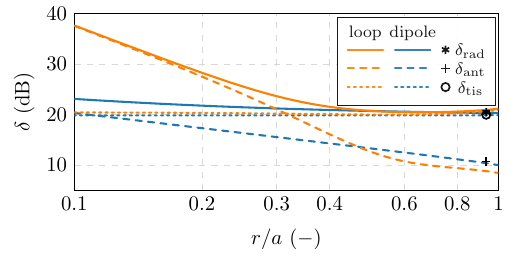}
    \caption{Dissipation factors of the meander dipole (black markers), a circular loop (orange lines), and a straight dipole antenna (blue lines). The same normalization as in Fig.~\ref{fig:effWater} is used.}
    \label{fig:disWaterDL}
\end{figure}

In order to understand the design principles of implanted antennas in more detail, it is worth studying the simple cases of loop and dipole antennas in Fig.~\ref{fig:effWater}. The dipole antenna has a positive trend in radiation efficiency with increasing length, while the loop antenna achieves its maximum performance around the radius of~$r = 0.6a$. These trends can be explained by separating dissipation factors into three contributions as depicted in Fig.~\ref{fig:disWaterDL}. With the increasing length of the dipole antenna, the tissue dissipation factor is constant, but on the contrary, the antenna dissipation factor declines, which is inherited by the total dissipation factor. The antenna dissipation factor of the circular loop antenna completely dominates for the small radius up to $r/a = 0.3$. Antenna dissipation factor decay is faster with the increasing radius for the circular loop than for the straight dipole antenna due to the impedance increase. However, with a circumscribing sphere radius of $0.5a$, the circumference of the circular loop antenna is comparable to the wavelength in the surrounding medium, and a further increase stops the decrease of the antenna dissipation factor and increase of tissue dissipation factor.

\begin{figure}
    \centering
    \includegraphics{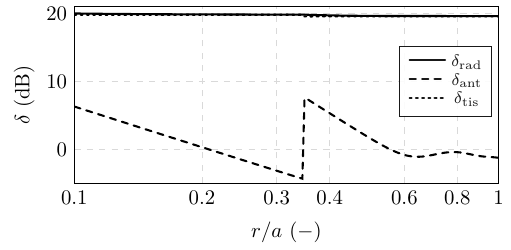}
    \caption{Dissipation factors corresponding to the upper bound on radiation efficiency for the water host body. The same normalization as in Fig.~\ref{fig:effWater} is used.}
    \label{fig:disWater}
\end{figure}
\begin{figure*}
    \centering
      \subfloat{\includegraphics[width = 1.66in]{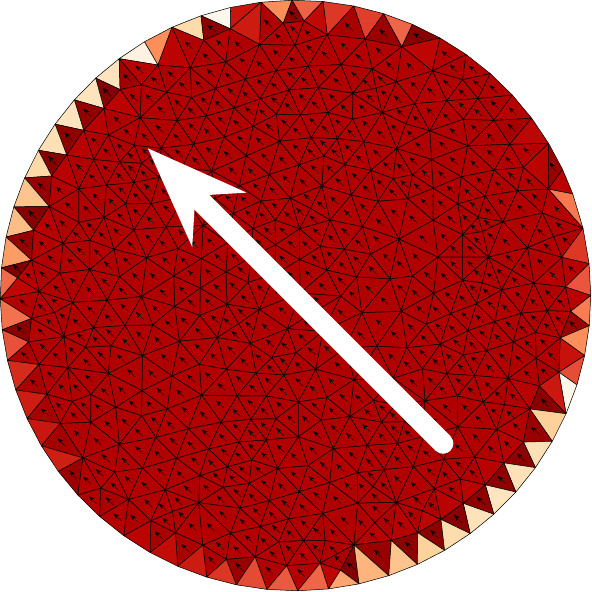}}
      \hfill
      \subfloat{\includegraphics[width = 1.66in]{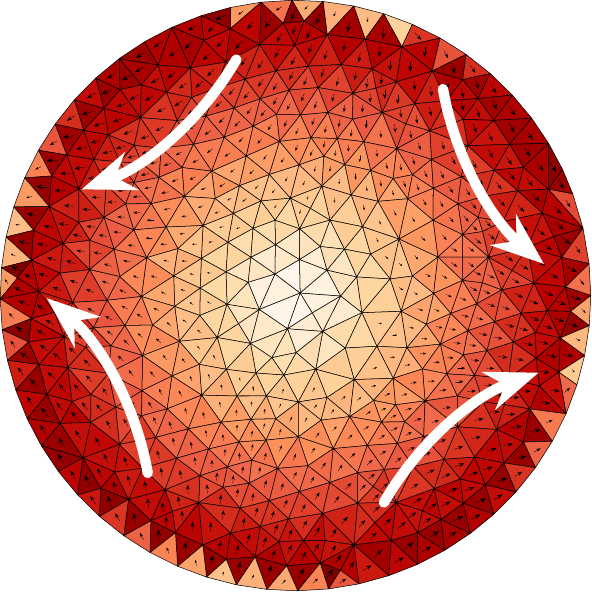}}
      \hfill
      \subfloat{\includegraphics[width = 1.66in]{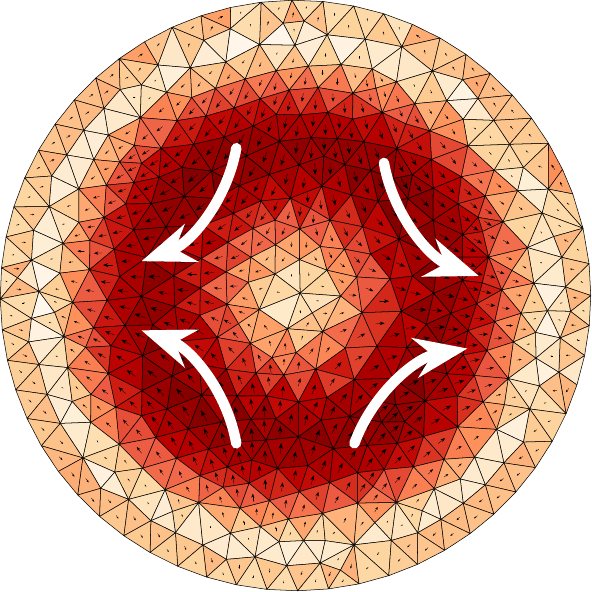}}
     \hfill
      \subfloat{\includegraphics[width = 1.66in]{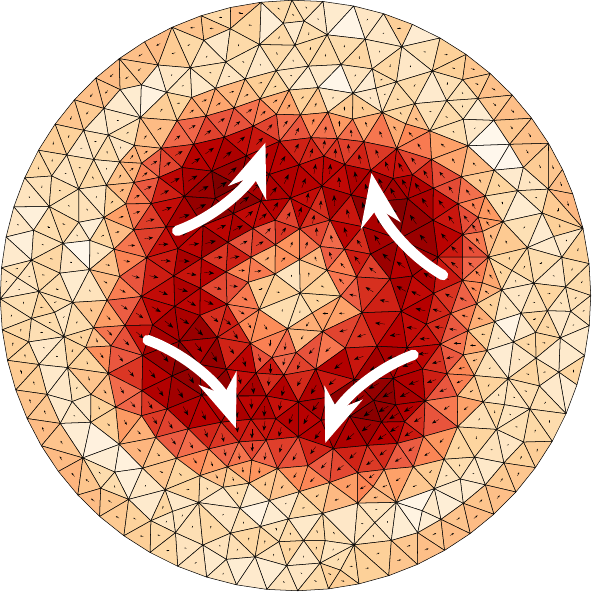}}
     \caption{Optimal current densities on a disc as the current-supporting region in the water phantom at $f_2$. The corresponding radii are (from left to right) $r/a = \{0.1, 0.4, 0.8, 1 \}$. The current densities are (from left to right) $\{\T{TM}_{1m}, \T{TM}_{2m}, \T{TM}_{2m}, \T{TM}_{2m} \}$-like.}
     \label{fig:optCurrentsWater}
\end{figure*}
Due to high antenna losses, the performance gap (distance to fundamental bound) for small electrical sizes is significant for straight dipole antennas (being comparable to tissue losses) and even wider for loop antennas. To see the reason for this behavior, it is instructive to plot how the dissipation factors, see Fig.~\ref{fig:disWater}, corresponding to the fundamental bound, behave when changing the size of the current-supporting region. The total dissipation factor is approximately constant, and it is dominated by the tissue dissipation factor, while the antenna dissipation factor is discontinuous at $r/a \approx 0.35$. As can be seen in Fig.~\ref{fig:optCurrentsWater}, for small electric sizes, a $\T{TM}_{1m}$-like current realizes the optimal current density. In contrast, the $\T{TM}_{2m}$-like current is optimal at larger electric sizes.  A switch between these two current profiles is represented by an abrupt jump seen in~Fig.~\ref{fig:disWater}. This critical point at $r/a \approx 0.35$ happens because of a slight improvement in radiation efficiency: a small drop in tissue dissipation factor and an increase in antenna dissipation factor; the radiation on an implanted antenna must always be seen as a trade-off between the antenna and tissue dissipation factors. The $\T{TM}_{2m}$-like current shows that in certain scenarios, antennas using higher-order modes can perform better than the ones using only the dipole currents.

\section{Conclusion}
The use of the hybrid method for the computation of radiation efficiency limitations for implanted antennas is shown in this paper. The hybrid method not only allows for a dramatic reduction of computational demands (order in magnitude faster than commercial solutions of the same setup) but also reveals different contributions of losses in the problem of an antenna radiating through lossy media. The technique is applied to a disc-like current-supporting region that is encapsulated in a dielectric and placed in the host body, i.e., a setup able to accommodate many of the practical implanted antenna designs. The paper shows how the upper bound on the radiation efficiency of all these designs can simply be evaluated. Furthermore, the paper reveals the importance of ohmic loss in the antenna proper compared to tissue loss.

Varying the size of the current-supporting region, the trade-off between ohmic losses in the antenna and the tissue drastically changes the shape of the optimal current density. The critical points were identified where the $\T{TE}_{1m}$-like current density performs better than a $\T{TM}_{1m}$-like current and vice-versa. This proves the importance of the fundamental limitations on the performance of implanted antennas and the usefulness of evaluating them for the scenario before starting the design to know the bounds.

The method is verified by measurement and by computation in commercial solvers, with the example being a meander dipole with a balun immersed in a phantom filled with distilled water. Great agreement is achieved for the results from the hybrid method, commercial simulators, and measurement. A detailed study of the fundamental bound on radiation efficiency in this scenario has also been done, concluding that ohmic loss in an implanted antenna must be considered in the design.


%

\appendices
\section{Method of Moments and T-matrix Hybrid} \label{app:hybrid}
\Ac{MoM} and T-matrix methods are described in~\cite{2022_Losenicky_TAP}, and only the essential steps to hybridize them are presented here. As mentioned above, the method applies \ac{MoM} to the antenna description and \ac{VSWE} with T-matrix for the surrounding tissues.

\subsection{Method of Moments}
Within \ac{MoM}, the current-supporting region, see~$\varOmega_Z$ region in Fig.~\ref{fig:setupLayer}, is discretized in a set of elementary cells together with current density, which is approximated by a sum of basis functions~$\left\{\basisFcn_n\right\}_{n=1}^N$ weighted by coefficients~$I_n$,
\begin{equation} \label{eq:currentExpansion}
    \V{J} (\V{r}) \approx \sum \limits_{n=1}^N I_n \basisFcn_n (\V{r}).
\end{equation}
Substitution of~\eqref{eq:currentExpansion} in the \ac{EFIE} and the use of Galerkin method~\cite{Harrington_FieldComputationByMoM} results in a system of linear equations
\begin{equation} \label{eq:MoM}
\left( \Zmat_0 + \Zmat_\rho \right) \Ivec = \Vvec_\T{i},
\end{equation}
where $\Zmat_0$ is the free-space impedance matrix, $\Zmat_\rho$ is the material impedance matrix, $\Vvec_\T{i}$ is the excitation vector, and $\Ivec$ is the unknown vector of current expansion coefficients~$I_n$. This allows for the characterization of the antenna in the free space. It should also be noted that the addition of dielectric inclusions or the use of non-spherical encapsulation can be resolved by the use of a volumetric electric field integral equation in the same way as described above.

\subsection{Vector Spherical Wave Expansion}
The field expansion into a set of basis functions is also applied to the T-matrix method. In this case, entire-domain basis functions are used, and the electric and magnetic fields are expanded in vector spherical harmonics
\begin{align}
    \V{E} (\V{r}) &= k \sqrt{Z} \sum \limits_{\alpha} a_\alpha \M{u}_\alpha^{(1)} (k \V{r}) + f_\alpha \M{u}_\alpha^{(4)} (k \V{r}), \\
    \V{H} (\V{r})&= \J \dfrac{k}{\sqrt{Z}} \sum \limits_{\alpha} a_\alpha \M{u}_{\bar{\alpha}}^{(1)} (k \V{r}) + f_\alpha \M{u}_{\bar{\alpha}}^{(4)} (k \V{r}),
\end{align}
where $k$ and $Z$ are wave number and background wave impedance, respectively, $a_\alpha,~f_\alpha$ are the expansion coefficients, and $\M{u}_\alpha^{(1)},~\M{u}_\alpha^{(4)}$ are regular and outgoing \acp{VSW}~\cite[Appendix~A]{2022_Losenicky_TAP}, respectively. To describe the scenario of an implanted antenna, the relation between the field outside the body and in the implant (antenna) encapsulation is given in matrix form as
\begin{equation}\label{eq:VSWE}
    \begin{bmatrix}
        \M{f}_1 \\ \M{a}_2
    \end{bmatrix} = \begin{bmatrix}
        \M{T} & \M{\Psi} \\
        \M{\Psi}^\trans & \M{\Gamma}
    \end{bmatrix} \begin{bmatrix}
        \M{a}_1 \\ \M{f}_2
    \end{bmatrix},
\end{equation}
with $\M{T},~\M{\Gamma}$ being transmission matrices outside and inside, respectively, matrices $M{\Psi},~\M{\Psi}^\trans$ accounting for field penetrating from inside to outside or vice-versa, and $\M{a},~\M{f}$ gathering expansion coefficients. Sub-indices ``1'' and ``2'' correspond to expansion outside and inside, respectively; see Fig.~\ref{fig:setupLayer}. In this paper, neither an incident field from infinity nor an external antenna is considered. Therefore, $\M{a}_1 = \M{0}$ is used for simplicity in further derivations.

In the case of a spherically layered host body, the interaction matrices in~\eqref{eq:VSWE} can be evaluated analytically~\cite{Merli-TheEffectOfInsulatingLayersOnThePerformanceOfImplantedAntennas}, which is also done in this paper. A representative example is a spherical encapsulation with an antenna centered in a phantom of the human head. The human head can be sufficiently approximated by a spherical multilayered host body~\cite{Kim-ImplantedAntennasInsideHumanBody} where tighter bounds on implanted antennas, as compared to~\cite{Skrivervik_Bosiljevac_Sipus_FundamentalBoundsForImplantedAntennas,Gao_OnTheMaximumPowerDensityofImplantedAntennasWithinSimplifiedBodyPhantomsEuCAP22}, can be evaluated in terms of seconds using the technique proposed here. The method can be easily adapted to other scenarios, such as ingestible devices and is not restricted to medical telemetry, but the matrices~\eqref{eq:VSWE} have to be obtained numerically in the case of a non-spherical model.

\subsection{Hybridization}
The coupling of the \ac{MoM} and T-matrix method is done via projection matrix~$\M{U}_1$ defined element-wise as
\begin{equation}
    U_1^{\alpha n} = k \sqrt{Z} \left\langle \M{u}_\alpha^{(1)}, \basisFcn_n \right\rangle,
\end{equation}
which maps \acp{VSW} to basis functions from \ac{MoM} as
\begin{align}
    \M{f}_2 &= - \M{U}_1 \Ivec, \label{eq:fUI} \\
    \Vvec &= \Vvec_\T{i} + \M{U}_1^\trans \M{a}_2. \label{eq:VVUa}
\end{align}
The above relations are written for a particular case of outgoing \acp{VSW} and complete excitation within \ac{MoM}, where $\Vvec_\T{i}$ is the direct excitation in the current-supporting region (for example, the delta-gap source).

Integrating both methods necessitates using a distinct spherical boundary. This characteristic is integral to the approach for calculating performance limitations. Inside the spherical boundary, the volume is treated by \ac{MoM} and outside by \ac{VSWE}. This work considers the spherical lossless encapsulation with the surface current-supporting region centered inside the spherical host body. This allows the use of 2D-\ac{MoM} for the current-supporting region inside the lossless encapsulation and the analytic computation of \ac{VSWE} matrices of a spherical host body. The computation demands are presented in Appendix~\ref{app:computation}. When current support is placed inside an arbitrarily shaped encapsulation, a combination of surface and volumetric \ac{MoM} must be used within the spherical boundary. Also, once the host body is not of spherical geometry, a numerical evaluation of matrices coupling \acp{VSW} is necessary. Such computation can be performed via sequential illuminating of the host body from inside and outside by \acp{VSW} and storing the response in the corresponding matrices.

\section{Computation Requirements}\label{app:computation}
Computation times and memory requirements are discussed in this appendix. The evaluations in \ac{AToM} and \ac{CST} are performed on computer using an Intel(R) Core(TM) i7-10700 CPU, 32.0\,GB RAM and 
an NVIDIA T1000 (3718\,MB) GPU.

Computation in \ac{AToM} is based on hybrid method of \ac{MoM} and \ac{VSWE} and subsequent optimization using \ac{QCQP}. Within \ac{MoM}, the considered disc current-supporting region is represented by 1026 Rao-Wilton-Glison functions~\cite{RaoWiltonGlisson_ElectromagneticScatteringBySurfacesOfArbitraryShape}. \ac{MoM} matrices are evaluated with the help of the GPU. Following~\cite{2022_Losenicky_TAP}, less than four hundred different \acp{VSW} represent the host body in all scenarios. The number of degrees of freedom in the optimization problem is equal to the amount of \ac{MoM} basis functions. The \ac{QCQP} is solved as \ac{GEP}, see Appendix~\ref{app:optim}. Overall memory requirements do not exceed 100\,MB for evaluating single current-supporting size in Figs.~\ref{fig:effMuscle},~\ref{fig:disMuscle} or~\ref{fig:effWater},~\ref{fig:disWater}. Computation times needed to evaluate principal limitations, using the hybrid approach within \ac{AToM} for a single current-supporting region's size are presented in Table~\ref{tab:ctHyb}.
\begin{table}
\centering
\caption{Computation times (in seconds) for the evaluation of principal limitations for different scenarios. ``muscle'': limitation at frequency~$f_1$ and current-supporting region's size $r/a=1$ inside a spherical air bubble within the host body as described in Section~\ref{sec:losses}. ``water'': limitation at frequency~$f_2$ and current-supporting region's size $r/a=1$ inside a spherical ceramic encapsulation within the host body as described in Section~\ref{sec:verification}. The most computationally demanding representative example is chosen, \ie, the highest frequency and the largest size of the current-supporting region.}
 \begin{tabular}{c c c c c} 
 scenario \textbackslash{} technique & \ac{MoM} & \ac{VSWE} & \ac{QCQP} & total \\
 \hline
 ``muscle'' & 2.40\,s & 0.28\,s & 0.12\,s & 3.44\,s \\
 ``water'' & 3.15\,s & 0.46\,s & 0.13\,s & 4.78\,s \\
 \hline
 \end{tabular}
 \label{tab:ctHyb}
\end{table}
The times demonstrate the low computation complexity of the approach used. The difference between the two scenarios shown in Table~\ref{tab:ctHyb} is mainly caused by different wavelengths inside the encapsulation and, therefore, different amounts of considered \acp{VSW}. Most of the time is spent in the computation of \ac{MoM} matrices.

To compare the time and memory requirements of different solvers for the simulation of the set-up from Fig.~\ref{fig:effWater}, and to relate the times of computation to the evaluation of principal limitations, the requirements of the solvers, \ac{AToM}, \ac{CST}, and \ac{FEKO}, are shown in Table~\ref{tab:comp}.
\begin{table}
\centering
 \caption{Computation times and memory usage of simulations in different solvers and approaches: hybrid method within \ac{AToM}, \ac{FDTD} within \ac{CST}, and \ac{MoM} with surface equivalence within \ac{FEKO}. The setup from Fig.~\ref{fig:gainPattern} is considered.}
 \begin{tabular}{c | c c c} 
 resouce & \ac{AToM} & \ac{CST} & \ac{FEKO} \\
 \hline
 memory (MB) & $87$ & $5,984$ & $42,606$ \\
 time (s) & $13$ & $5,655$ & $15,840$
 \end{tabular}
 \label{tab:comp}
\end{table}
The computations in \ac{AToM} and \ac{CST} are performed using the same computer, while computation in \ac{FEKO} is more demanding, a machine with the following parameters is used: two Intel Xeon E5-2665 CPUs and 72\,GB RAM. In the approach via the hybrid method using \ac{AToM}, the meander dipole antenna is represented by 805 Rao-Wilton-Glisson basis functions~\cite{RaoWiltonGlisson_ElectromagneticScatteringBySurfacesOfArbitraryShape} in \ac{MoM} and the host body by 336 \acp{VSW}. In \ac{CST}, the entire model is discretized in $102,648,00$~mesh cells\footnote{The \ac{PEC} symmetry plane is applied to reduce the number of cells and speed up the computation.} and 72,160 time steps in the used \ac{FDTD} solver. In \ac{FEKO}, the model is represented using the surface equivalence principle with 147450~basis functions. Calculation of matrix elements took $8,166\,\T{s}$, solution of the system of linear equations $7,631\,\T{s}$ and its preconditioning $21\,\T{s}$. Computation via a hybrid approach has extremely low memory and time computation requirements compared to others, as shown in Table~\ref{tab:comp}.

\section{Optimization}\label{app:optim}
As seen in~\eqref{eq:Prad}--\eqref{eq:Ptis}, all the power terms are represented as quadratic functions of expansion coefficients. Converting all terms~\eqref{eq:Prad}--\eqref{eq:Ptis} to the functionals of vectors~$\Ivec$ via~\eqref{eq:linCon}, the optimization problem~\eqref{eq:optim} is transformed to \ac{QCQP}. Its solution can be approached numerically~\cite{Liska_etal_FundamentalBoundsEvaluation}, but it can be solved more efficiently in this particular case. After several algebraic manipulations and substitutions from~\eqref{eq:linCon} to~\eqref{eq:Prad}--\eqref{eq:Ptis}, the optimization problem~\eqref{eq:optim} reduces to \ac{GEP}
\begin{equation}
\left( \M{U}_1^\trans \M{\Psi}^\trans \M{\Psi} \M{U}_1 \right) \Ivec = \eta \left(\RE \left[\Zmat_0 +  \Zmat_\rho \right] + \M{U}_1^\trans \M{\Gamma} \M{U}_1 \right) \Ivec,
\end{equation}
where the left-hand-side matrix is proportional to the cycle-mean radiated power~\eqref{eq:Prad} and the right-hand side to the total cycle-mean supplied power. The maximum radiation efficiency~$\eta_\T{rad}^\T{ub}$ is given by the maximum eigenvalue~$\eta$, and the optimal current density corresponds to the associated eigenvector~$\Ivec$. As only one eigenvalue is needed, the \ac{GEP} solution time is negligible, see Appendix~\ref{app:computation}.

\section{Measurement} \label{app:meas}
The implanted meander dipole antenna within a glass bottle filled with distilled water gain pattern is measured at a distance of 2.1\,m from the reference antenna (a quad-ridge horn antenna, QH400, MVG~Industries), see Fig.~\ref{fig:photo}. The meander dipole antenna is oriented along the $y$ axis.

\begin{figure}
     \centering
     \includegraphics{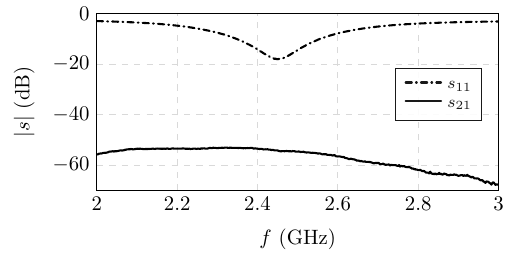}
     \caption{Measured magnitudes of reflection and transmission coefficients of the implanted meander dipole antenna within a glass bottle filled with distilled water; performed in the anechoic chamber.}
     \label{fig:mS11S12}
 \end{figure}

The reflection and transmission coefficients were measured from 2\,GHz to 3\,GHz, see Fig.~\ref{fig:mS11S12}, evidencing good matching of the meander dipole antenna at frequency~$f_2$.
 At studied frequency~$f_2$, measured magnitudes of reflection and transmission coefficients are $|s_{11}| = -18 \, \T{dB}$ and $|s_{12}| = -54 \, \T{dB}$, respectively. The transmission coefficient was measured in the direction of gain pattern maximum $\theta = \pi/2$ and $\varphi = 0$. In all measurements, the vector network analyzer was calibrated to the reference plane lying at coaxial ports placed at the end of a semi-rigid cable feeding the antenna.

The gain pattern of the implanted meander dipole antenna, $G(\theta, \varphi) = \eta_\T{rad} D(\theta, \varphi)$, is obtained from the measured transmission coefficient by excluding the gain of the reference antenna and the path loss between the two antennas, which was calibrated as $-36.10$\,dB in total at frequency~$f_2$ in the measurement setup. The measured gain pattern was further corrected by the balun insertion loss of 0.30\,dB and the cable loss of 0.35\,dB (cable connected from the measurement reference plane to the balun).


\ifCLASSOPTIONcaptionsoff
  \newpage
\fi



\bibliographystyle{IEEEtran}
\bibliography{references.bib}

\begin{IEEEbiography}[{\includegraphics[width=1in,height=1.25in,clip,keepaspectratio]{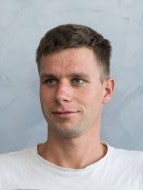}}]{Jakub Liska}
received his B.Sc. and M.Sc. degrees in electrical engineering from the Czech Technical University in Prague, Czech Republic, in 2019 and 2021, respectively. He is currently continuing towards a Ph.D. degree in electrical engineering at the same university.

His research interests include \ac{EM} field theory, fundamental bounds, computational electromagnetics, numerical and convex optimization, numerical techniques, and eigenproblems.
\end{IEEEbiography}

\begin{IEEEbiography}[{\includegraphics[width=1in,height=1.25in,clip,keepaspectratio]{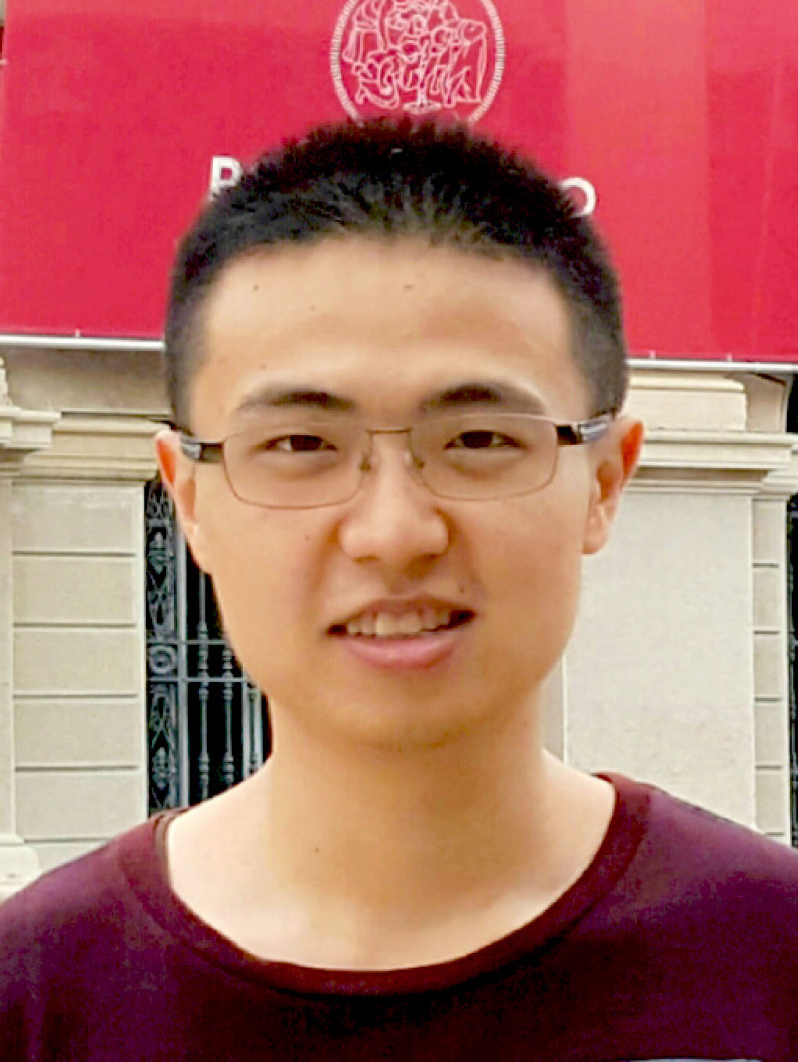}}]{Mingxiang Gao}
received the double B.Sc. degree in electrical engineering and business administration and the M.Sc. degree in electrical engineering from the Xi’an Jiaotong University, Xi’an, China, in 2016 and 2019, respectively, and the M.Sc. degree (summa cum laude) in electrical engineering from the Politecnico di Milano, Milan, Italy, in 2019. He is currently working towards the Ph.D. degree in electrical engineering at the École Polytechnique Fédérale de Lausanne, Lausanne, Switzerland. His current research interests include the theory and design of implantable antennas, wireless techniques for bioelectronics, and bioelectromagnetics.
\end{IEEEbiography}

\begin{IEEEbiography}[{\includegraphics[width=1in,height=1.25in,clip,keepaspectratio]{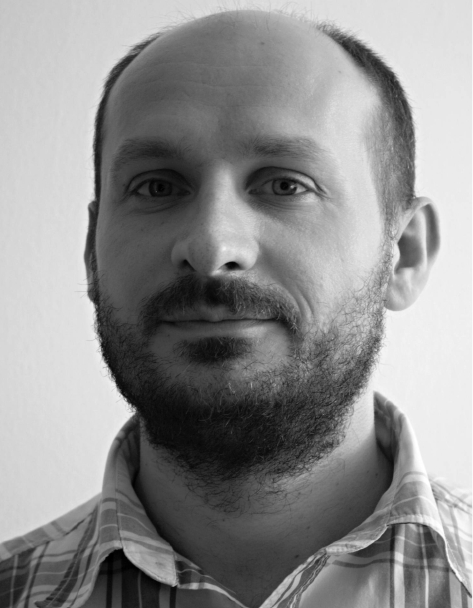}}]{Lukas Jelinek}
received his Ph.D. degree from the Czech Technical University in Prague, Czech Republic, in 2006. In 2015 he was appointed Associate Professor at the Department of Electromagnetic Field at the same university.

His research interests include wave propagation in complex media, electromagnetic field theory, metamaterials, numerical techniques, and optimization.
\end{IEEEbiography}

\begin{IEEEbiography}[{\includegraphics[width=1in,height=1.25in,clip,keepaspectratio]{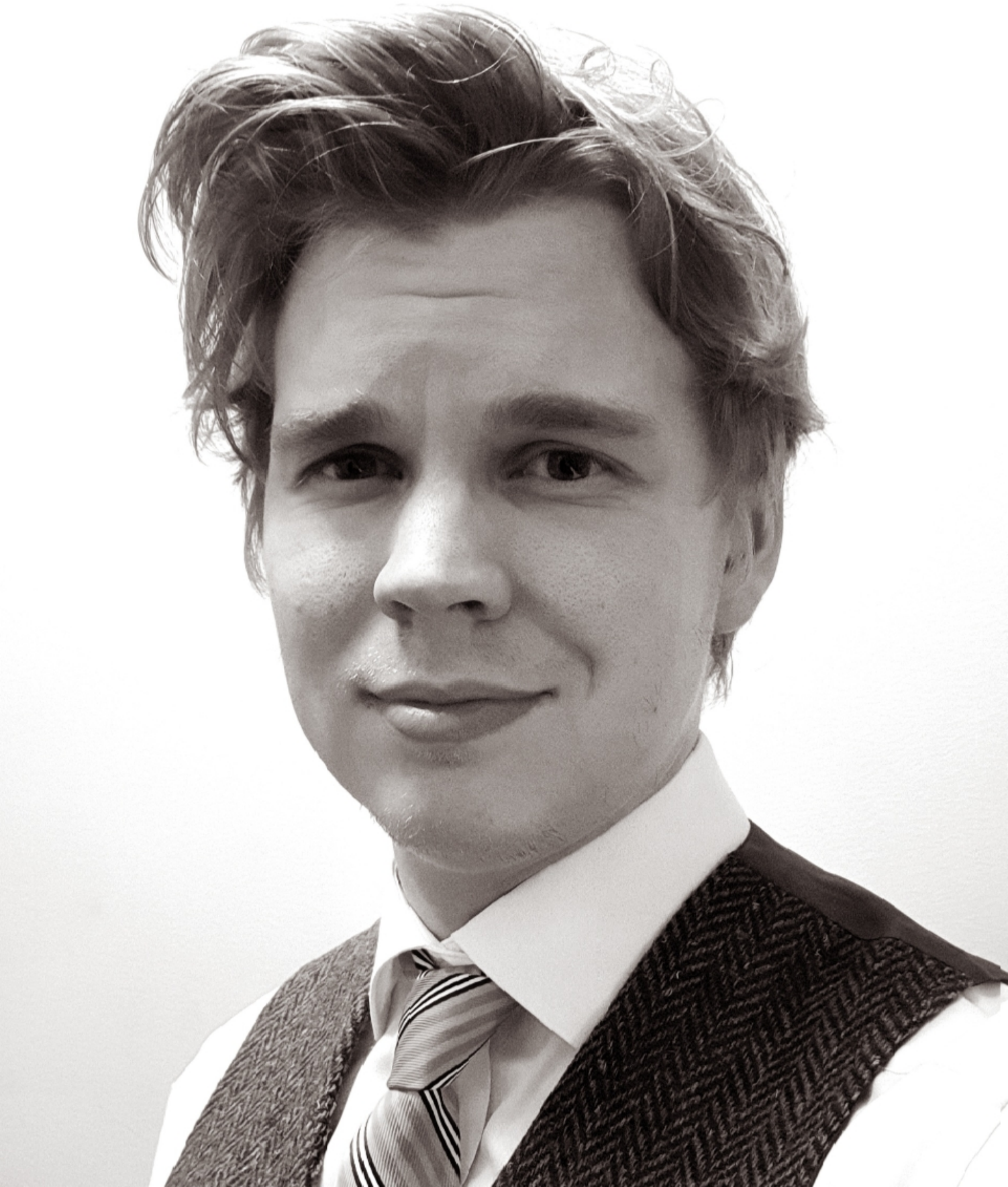}}]{Erik R. Algarp} received his M.Sc. degree in Electrical Engineering from KTH Royal Institute of Technology, Sweden, in 2022. He is currently pursuing a Ph.D degree in Electrical Engineering within the Microwaves and Antennas group at École polytechnique fédérale de Lausanne. 
\end{IEEEbiography}

\begin{IEEEbiography}[{\includegraphics[width=1in,height=1.25in,clip,keepaspectratio]{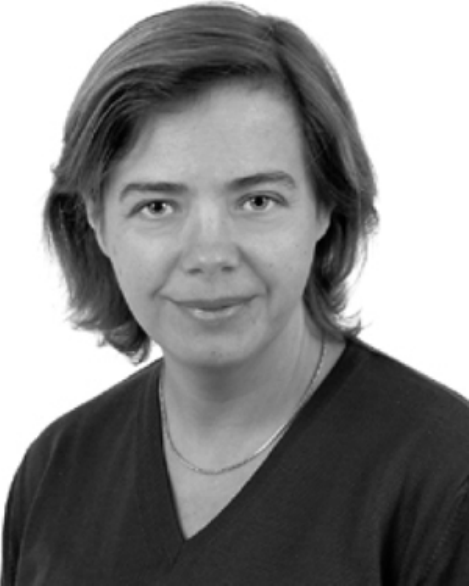}}]{Anja K. Skrivervik} received the master’s and Ph.D. degrees in electrical engineering from the École Polytechnique Fédérale de Lausanne (EPFL), Lausanne, Switzerland, in 1986 and 1992, respectively.

After a stay at the University of Rennes, Rennes, France, as an Invited Research Fellow and two years in the industry, she returned part-time to EPFL as an Assistant Professor in 1996, where she is currently a Professeur Titulaire, the Head of the Microwave and Antenna Group. She was the Director of the EE Section from 1996 to 2000, and she is currently the Director of the EE Doctoral School at EPFL. She is also a Visiting Professor at the University of Lund. Her teaching activities include courses on microwaves and antennas, and she teaches at bachelor’s, master’s, and Ph.D. levels. She is the author or co-author of more than 200 peer-reviewed scientific publications. Her research interests include electrically small antennas, antennas in biological media, periodic structures, reflect and transmit arrays, and numerical techniques for electromagnetics.

Dr. Skrivervik is a Board Member of the European School on Antennas and is frequently requested to review research programs and centers in Europe. She has been a member of the Board of Directors of the European Association on Antennas and Propagation (EurAAP) since 2017. She received the Latsis Award. She is very active in European collaboration and European projects. She was the Chairperson of the Swiss URSI until 2012. She was the General Chair of the Loughborough Antenna and Propagation Conference in 2015, the Vice-Chair and Technical Program Committee-Chair of the EuCAP 2016 Conference, and the Financial Chair of EuCAP 2017 to EuCAP 2022. 
\end{IEEEbiography}

\begin{IEEEbiography}[{\includegraphics[width=1in,height=1.25in,clip,keepaspectratio]{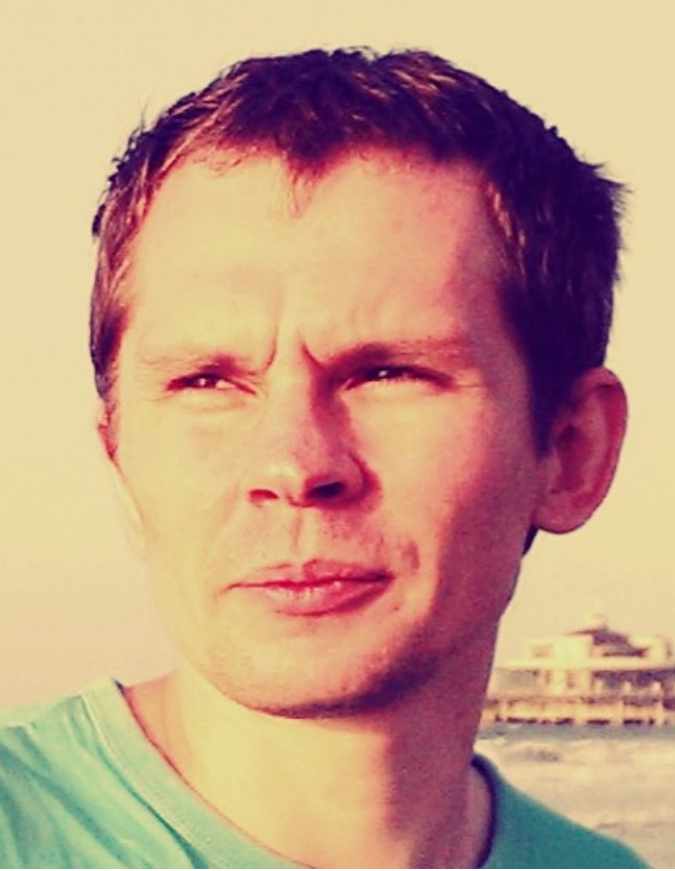}}]{Miloslav Capek}
(M'14, SM'17) received the M.Sc. degree in Electrical Engineering 2009, the Ph.D. degree in 2014, and was appointed a Full Professor in 2023, all from the Czech Technical University in Prague, Czech Republic.
	
He leads the development of the AToM (Antenna Toolbox for Matlab) package. His research interests include electromagnetic theory, electrically small antennas, antenna design, numerical techniques, and optimization. He authored or co-authored over 160~journal and conference papers.

Dr. Capek is the Associate Editor of IET Microwaves, Antennas \& Propagation. He was a regional delegate of EurAAP between 2015 and 2020 and an associate editor of Radioengineering between 2015 and 2018. He received the IEEE Antennas and Propagation Edward E. Altshuler Prize Paper Award~2023.
\end{IEEEbiography}

\end{document}